\documentclass[10pt,a4paper,aps,twocolumn,showpacs,superscriptaddress,nofootinbib]{revtex4-1}
\usepackage[colorlinks,hyperindex]{hyperref}
\usepackage{epsfig}
\usepackage{float}
\RequirePackage{amssymb,amsmath}
\usepackage[latin1]{inputenc}
\usepackage{graphicx}
\usepackage{indentfirst}
\usepackage{color}
\begin{document}
\title{Configurational entropy and $\rho$ and $\phi$ mesons production in QCD}
\author{G. Karapetyan}
\email{gayane.karapetyan@ufabc.edu.br}
\affiliation{Centro de Ci\^encias Naturais e Humanas, Universidade Federal do ABC - UFABC, 09210-580, Santo Andr\'e, Brazil}
\begin{abstract}
In the present work the electroproduction for diffractive $\rho$ and $\phi$ mesons by considering AdS/QCD correspondence and Color Glass Condensate (CGC) approximation are studied with respect to the associated dipole cross section, whose parameters are  studied and analysed in the framework  of the configurational entropy. Our results suggest different quantum states of the nuclear matter, showing that the extremal points of the nuclear configurational entropy is able to reflect a true description of the $\rho$ and $\phi$ mesons production, using current data concerning light quark masses.
During the computations parameters, obtained in fitting procedure, coincide to the experimental within $\sim0.1\%$. 
\end{abstract}
\maketitle
\section{Introduction}
Recent studies concerning lattice QCD setup and the AdS/QCD approximation  have shown the importance of the configurational entropy scheme in the domain of high energy and particle nuclear physics. Important aspects of high energy nuclear and particle physics can be derived from configurational entropic setup. 
The lower experimental occurrence of mesons of higher spin and was  quantitatively derived in Ref. \cite{Bernardini:2016hvx}, whereas the dominance of lower spin glueball states  was scrutinized in Ref. \cite{Bernardini:2016qit}, using the configurational entropy. Besides, 
Bose--Einstein condensates critical densities were derived from the relative configurational entropy critical points \cite{Casadio:2016aum}, while the holographic model for quarkonia was studied in this framework in Ref. \cite{Braga:2017fsb}, showing the lower prevalence of higher spin charmonium and bottomonium states. Hawking--Page transitions were derived 
as a byproduct of the critical points of the configurational entropy in Ref. \cite{Braga:2016wzx} and KdV solitons through a cold quark-gluon plasma were investigated in Ref. \cite{daSilva:2017jay}.
The configurational entropy can comprise all the possible spatially localized nuclear configurations that have relevant data regarding  strong interactions in QCD \cite{Ma:2018wtw}, providing a new observable to analyze a nuclear system evolution.

 The configurational entropy takes into account the localized energy densities  \cite{Gleiser:2012tu,Sowinski:2015cfa,Gleiser:2011di,Gleiser:2015rwa,Gleiser:2013mga}. When cross sections are taken into account, instead, the nuclear configurational entropy was shown to be a relevant paradigm to study nuclear and particle physics 
\cite{Karapetyan:2017edu,Karapetyan:2016fai}, also based on the Fourier transformation of functions that depend on the position and are spatially confined, as the involved cross sections \cite{Alves:2017ljt,Alves:2016koo}.
The cross section of the particle/fragment production as well as the probability of the different reactions in nuclear physics can be obtained and estimated using configurational entropy techniques. The cross section, as a probability of that a nuclear reaction will occur, defines a characteristic area of the interaction, and as a spatially confined function, it can be used in the frame of the configurational entropy. 

On the other hand, anti-de Sitter (AdS) spacetime together is the stage for a dual prescription of quantum-chromodynamics (QCD) can be considered as one of the most progressive approaches for the study of the structure and nuclear properties of hadrons and their interactions, being a state-of-the-art in the field of non-perturbative QCD \cite{Karch:2006pv}.
Such techniques allow to investigate, for instance, the form factors and the transverse charge densities of the deuteron, considering the unpolarised and the transversely polarised cases \cite{Huang}, and show a good agreement with the phenomenological parametrisation and experimental data. 
The holographic light-front wavefunction for the diffractive $\rho$ and  $\phi$ electroproduction, in the frame of AdS/QCD correspondence and the Color Glass Condensate (CGC), was used to predict the dipole cross section, and the parameters, obtained during the fitting procedure, in full compliance with the most recent 2015 high precision HERA data on inclusive Deep Inelastic Scattering  \cite{Ahmady}.
The interest in the electroproduction of vector mesons is multifold, from the theoretical to the experimental points of view. 
Experiments at HERA demonstrated the diffractive scattering at higher energies.
From the theoretical point of view,  the exclusive diffractive production of vector mesons in DIS is an absolutely hard phenomenon, and the main features of it can be explained by QCD theory.
Indeed, there is a huge amount of problems and applicative tasks, which could be solved and interpreted using these techniques, which will match of obtained experimental and phenomenological results. This article is devoted to the investigation of parameters of diffractive $\rho$ and $\phi$ production, in the context of the AdS/QCD and the Color Glass Condensate, using information entropy.

The present paper is divided as follows: Sect. II is devoted to give some details of the AdS/QCD and the Color Glass Condensate (CGC) regime approximation. We also present the computation of the points of minima of the configurational entropy, in order to predict 
the parameters for diffractive $\rho$ and $\phi$ vector mesons production. We conclude in Sect. III, also providing perspectives.

\section{Dipole model and configurational entropy}
QCD color dipole models and non-perturbative holographic meson light-front wavefunctions have been used to predict the cross sections for diffractive $\rho$ and $\phi$ electroproduction, measured at the HERA collider \cite{Ahmady}.
The dipole model consists of an effective theory describing high energy scattering in QCD, where the dipole cross section for the diffractive process $\gamma^* p \to V p$ is given by the following: 
\begin{eqnarray}
 \mathcal{A}_\lambda(t,s;Q^2)  
 &=& \sum_{h, \bar{h}} \int {\mathrm d}^2 {\mathbf r} \; {\mathrm d} x \; \Psi^{\gamma^*,\lambda}_{h, \bar{h}}(r,x;Q^2)  \Psi^{V,\lambda}_{h, \bar{h}}(r,x)
  \,\nonumber\\ 
&&\times \exp({-i x \mathrm{r} \cdot \Delta})\mathcal{N}(x_{{m}},\mathrm{r}, \Delta),
\label{amplitude-VMP} 
\end{eqnarray}
where $s$ is the squared centre-of-mass energy, $t=-\Delta^2$ is the squared momentum transfer at the proton vertex. The index $\lambda$ refers to the polarisation of the photon or the vector meson, $\lambda = L\;(\text{\rm longitudinal}), T\; (\rm transversal)$, $\Psi^{\gamma^*,\lambda}_{h, \bar{h}}(r,x;Q^2)$ and $\Psi^{V,\lambda}_{h, \bar{h}}(r,x)$  are the light-front wavefunctions of the photon and vector meson, respectively. Besides, $h$ is the helicity of the quark and $\bar{h}$ is a helicity of the antiquark in a $q\bar{q}$ color dipole of the transverse size $r$, and $x$ is a fraction of light-front momentum of the photon (or vector meson) carried by the quark, $\mathcal{N}(x_{{m}},\mathrm{r},\mathrm{\Delta})$ is a proton-dipole scattering amplitude, which depends on the modified Bjorken variable $x_{{m}}$, where
\begin{equation}
	x_{{m}}=x_{\text{Bj}}\left(1+ \frac{M_V^2}{Q^2} \right),
\label{Bjorken-x}
\end{equation}
where $x_{\text{Bj}}=\frac{Q^2}{W^2}$. 
In Eq. (2), $M_V$ is the mass of a vector meson, $Q^2$ is the photon virtuality and $W$ is an invariant mass of the final state hadronic system. 

If one substitutes the vector meson by a virtual photon in Eq. \eqref{amplitude-VMP}, one obtains the amplitude for elastic Compton scattering $\gamma^* p \to \gamma^* p$:
\begin{eqnarray}
\!\!\!\!\!\!\!\!\!\!\!\!\left. \mathcal{A}_\lambda(s,t) \right|_{t=0}  
 \!&\!=\!&\!  s \sum_{h, \bar{h}} \int {\mathrm d}^2 {\mathbf r} {\mathrm d} x \; |\Psi^{\gamma^*,\lambda}_{h, \bar{h}}(r,x;Q^2)|^2  \hat{\sigma}(x_{{m}}, r),
\label{amplitude-compton} 
\end{eqnarray}
where the dipole cross section is given by
\begin{equation}
	\hat{\sigma}(x_{{m}},r)=\frac{\mathcal{N}(x_{{m}},\mathrm{r}, \mathbf{0})}  {s}=\int \mathrm d^2 \mathbf{b}~\mathcal{\tilde{N}}(x_{{m}},\mathrm{r}, \mathbf{b}) \;.
	\label{dipole-xsec}
\end{equation}
The elastic amplitude given by Eq. \eqref{amplitude-compton} then can be transform using the optical theorem in order to obtain the inclusive $\gamma^* p \to X$ total cross section in DIS: 
\begin{equation}
\!\sigma_{\lambda}^{\gamma^* p \to X} \!=\! \sum_{h, \bar{h},f} \int {\mathrm d}^2 {\mathbf r} {\mathrm d} x  |\Psi^{\gamma^*,\lambda}_{h, \bar{h}}(r,x;Q^2)|^2  \hat{\sigma}(x_{{m}}, r).
\label{gammapxsec}
\end{equation}

By using  Eq. \eqref{gammapxsec} data for Deep Inelastic Scattering (DIS), it is possible to obtain the free parameters of the dipole cross section section and then use the same dipole cross section to make predictions for $\rho$ and  $\phi$ vector mesons production.
In order to estimate the total reaction cross section, one needs to calculate the differential cross section in the forward limit:
\begin{equation}
\left. {{\mathrm d} \sigma_\lambda \over dt}\right. \mid_{t=0}  
= \frac{1}{16\pi} 
[ \mathcal{A}_\lambda(s, t=0)]^2
\label{dxsection}
\end{equation}
with exponential $t$-dependence:   
\begin{equation}
\left. {{\mathrm d} \sigma_\lambda \over dt}\right.
= \frac{1}{16\pi} 
[\mathcal{A}_\lambda(s, t=0)]^2  e^{-B_D t}
\label{dsigmadt}
\end{equation}
with the diffractive slope parameter $B_D$:
\begin{equation}
B_D = 14N \left(\frac{1~\mathrm{GeV}^2}{Q^2 + M_{V}^2}\right)^{0.2}+N,
\label{Bslope}
\end{equation}
where $N\approxeq0.55$ GeV$^{-2}$. 
Eq. \eqref{dsigmadt} can be altered by: 
\begin{equation}
\left. {{\mathrm d} \sigma_\lambda \over dt}\right.
= \frac{1}{16\pi} 
[ \mathcal{A}_\lambda(s, t=0)]^2 \; (1 + \beta_\lambda^2) e^{-B_D t}
\label{dsigmadt-re}
\end{equation}
where the parameter $\beta_\lambda$ (the ratio of the real to the imaginary parts of the amplitude) is given by: 
\begin{equation}
\beta_\lambda=\tan \left(1.7 \alpha_\lambda \right) ~~ ~  {\rm with }~~ ~  \alpha_\lambda=\frac{\partial \log |\mathcal{A}_\lambda|}{\partial \log  \left(1/x\right)} \label{logderivative} \,. 
\end{equation}
Then one can obtain the cross section of photo-production, integrating Eq. \eqref{dsigmadt} over $t$. In recent studies \cite{Ahmady}, the calculation of the total cross section $ \sigma = \sigma_T + 0.98 \sigma_L$  has been done and compared to the HERA data. 

The $\rho$ and  $\phi$ vector meson light-front wavefunctions appearing in Eq. \eqref{amplitude-VMP}
can be expressed in the following form:
\begin{equation}
\Psi^{V,L}_{h,\bar{h}}(r,x) =  0.5 \delta_{h,-\bar{h}}  \bigg[ 1 + 
{ m_{f}^{2} -  \nabla_r^{2}  \over x(1-x)M^2_{V} } \bigg] \Psi_L(r,x) \,.
\label{mesonL}
\end{equation}
and
\begin{eqnarray}
\Psi^{V, T}_{h,\bar{h}}(r,x) &=& \pm \bigg[  i \exp(\pm i\theta_{r})  ( x \delta_{h\pm,\bar{h}\mp} - (1-x)  \delta_{h\mp,\bar{h}\pm})  \partial_{r}\nonumber\\&&\qquad+ m_{f}\delta_{h\pm,\bar{h}\pm} \bigg] {\Psi_T(r,x) \over 2 x (1-x)}\,. 
\label{mesonT}
\end{eqnarray}
where 
$m_f$ is a non-zero quark mass, $ \epsilon^{2} = x(1-x)Q^{2} + m_{f}^{2}$ and $r \exp(i \theta_{r})$ is the complex notation for the transverse separation between the quark and anti-quark. 
The meson wavefunction in a semiclassical approximation of light-front QCD with massless quarks can be written as \cite{Brodsky:2006uqa}
\begin{equation}
	\Psi(\zeta, x, \phi)= \exp(iL\phi) \mathcal{X}(x) \frac{\phi (\zeta)}{\sqrt{2 \pi \zeta}} 
\label{mesonwf}
\end{equation}
where $\zeta=\sqrt{x(1-x)} r$ is the transverse separation between the quark and the antiquark at equal light-front time, and  the transverse wavefunction $\phi(\zeta)$ is a solution of the holographic light-front Schr\"odinger equation,
 \begin{equation}
 	\left(-\frac{d^2}{d\zeta^2} - \frac{1-4L^2}{4\zeta^2} + U(\zeta) \right) \phi(\zeta) = M^2 \phi(\zeta), 
 	\label{holograhicSE}
 \end{equation}
where $M$ is the mass of the meson and $U(\zeta)$ is the confining potential. 
One can make the substitutions $\zeta \mapsto z$, for $z$ being the fifth dimension of the dual AdS space, together with  $L^2 -(2-J)^2 \mapsto (mR)^2$,  where $R$ and $m$ are the radius of curvature and mass parameter of AdS space respectively.
The potential is given by
 \begin{equation}
 	U(z, J)= 0.5 \varphi^{\prime\prime}(z) + 0.25 \varphi^{\prime}(z)^2 + \left(\frac{2J-3}{4 z} \right)\varphi^{\prime} (z), 
 \end{equation}
 where $\varphi(z)$ is the dilaton field, and a $\kappa$ is a single mass scale in the quadratic dilaton field, which was fixed to fit the observed slopes of the Regge trajectories for the various meson families, $\kappa \approx 0.5$ GeV. 
 A quadratic dilaton ($\varphi(z)=\kappa^2 z^2$) profile results in a harmonic oscillator potential in physical spacetime:
 \begin{equation}
 	U(\zeta,J)= \kappa^4 \zeta^2 + \kappa^2 (J-1) \;.
 	\label{harmonic-LF}
 \end{equation}
 One can obtain the meson mass spectrum by solving the holographic Schr\"odinger equation:
 \begin{equation}
 	M^2= 4\kappa^2 \left(n+\frac{L+J}{2}\right)\;
 	\label{mass-Regge}
 \end{equation}
  with the corresponding eigenfunctions
 \begin{equation}
 	\phi_{n,L}(\zeta)= \kappa^{1+L} \sqrt{\frac{2 n !}{(n+L)!}} \zeta^{0.5+L} e^{{\frac{-\kappa^2 \zeta^2}{2}}} L_n^L(x^2 \zeta^2) \;.
 \label{phi-zeta}
 \end{equation}

In order to simplify the holographic wavefunction given by Eq. \eqref{mesonwf}, the longitudinal wavefunction $\mathcal{X}(x)$ must be determined as:
 \begin{equation}
 	\mathcal{X}(x)=\sqrt{x(1-x)}.
 \end{equation}
For the ground state mesons with $n=0, L=0$  ($J=L+S$) , Eq. \eqref{mesonwf} can be presented in the form:
 \begin{equation} 
\Psi (x,\zeta)= \frac{\kappa}{\sqrt{\pi}} \sqrt {x(1-x) }  e^{{-{\kappa^2 \zeta^2 \over 2}}}, 
\label{wavef}
 \end{equation}
and a two-dimensional Fourier transform to momentum space reads
 \begin{equation}
\tilde{\Psi} (x,k) \propto  \frac{1}{\sqrt {x(1-x)}}  e^{{-\frac{M^2_{q\bar{q}}}{2\kappa^2}}}
\label{wavefk}
 \end{equation}
where $M^2_{q\bar{q}}$ is an invariant mass of the $q\bar{q}$ pair,  expressed by $
	M^2_{q\bar{q}}=\frac{k^2}{x(1-x)}$. 
The invariant mass then is expressed for non-zero quark masses as:
\begin{equation}
	M^2_{q\bar{q}}=\frac{k^2 + m_f^2}{x(1-x)} \;.
	\label{Massqbarq}
\end{equation}
Putting Eq. \eqref{Massqbarq} in Eq. \eqref{wavefk}, one can get the Fourier transform to configuration space as:
 \begin{equation}
   \Psi_{\lambda} (x,\zeta) = \mathcal{N}_{\lambda} \sqrt{x (1-x)}  e^{{ -{ \kappa^2 \zeta^2  \over 2}}}
e^{{ -{m_f^2 \over 2 \kappa^2 x(1-x) }}}
\label{hwf}
 \end{equation}
where ${\mathcal N}_{\lambda}$ is a polarisation-dependent normalization constant. 

The cross section of the $\rho$ and  $\phi$ diffractive mesons production in connection with impact parameter dependence ($b$-dependence) can be derived in frame of CGC dipole model as:
\begin{equation} 
\hat{\sigma}(x_{{m}},r) = \sigma_0 \, {{ \mathcal N } (x_{{m}}, r Q_s, 0 ) }\,.
\end{equation}
with
\begin{widetext}
\begin{eqnarray}
{ \mathcal N } (x_{{m}},r Q_s, 0) = \begin{cases}&
{ \mathcal N}_0 \left ( { 0.5 r Q_s }\right)^ {2 \left [ \gamma_s + {\log  (2 / r Q_s) \over  \kappa \, \lambda \, {\mathrm ln} (1/x_{{m}}) }\right]}, \quad ~{\rm for } \quad~ r Q_s \leq 2 \\ 
           & { 1- e^{-{\mathcal A} \,\log^2 ( {\mathcal B} \, r Q_s)}, } ~~~~~~~~~{\rm for } ~~~~~ ~~~~ r Q_s > 2,\end{cases}
\end{eqnarray}
\end{widetext}
where $Q_s = (x_0/x_{{m}})^{0.5 \lambda}$ GeV is a saturation scale and the coefficients ${\mathcal A}$  and ${\mathcal B}$ are defined as:
\begin{equation}
{\mathcal A} =  - { ({ \mathcal N}_0 \gamma_s)^2  \over (1 - { \mathcal N}_0)^2 \,\log[1 - { \mathcal N}_0] }\,, ~~~~~~
{\mathcal B} = {0.5} (1 - { \mathcal N}_0)^{-{(1 - { \mathcal N}_0) \over { \mathcal N}_0 \gamma_s }}\,.
\end{equation}
The three sets of adjustive parameters, $\sigma_0$ (which characterizes the non-perturbative contribution), $\lambda$ (the polarisation parameter), $x_0$ (which is a parameter fixed by the initial conditions), and $\gamma_s$ (an anomalous dimension parameter),  have been determined by fitting procedure for different set of effective quark masses $m_{u,d,s}$
for diffractive $\rho$ and $\phi$ production \cite{Ahmady}.
Thus, according to Ref. \cite{Ahmady},  for the ``Fit A" the fitted parameters are 
 \begin{equation}\!\frac{\sigma_0}{{\rm mb}} \!=\! 26.3, \;\gamma_s \!=\! 0.741,\;  \lambda \!=\! 0.219,\; x_0 \!=\! 1.81 \times10^{-5},\end{equation} with a $\chi^2/\mbox{d.o.f}=1.03$ in the case of effective quark masses $m_{u,d}, m_{s}=[0.046, 0.357]$ GeV.
 For the ``Fit B",  the authors chose following parameters: 
 \begin{equation}\!\frac{\sigma_0}{{\rm mb}} \!=\! 24.9,\;\; \gamma_s \!=\! 0.722,\;  \lambda \!=\! 0.222,\; x_0 \!=\! 1.80 \times10^{-5},\end{equation} with a $\chi^2/\mbox{d.o.f}=1.02$ in the case of effective quark masses $m_{u,d}, m_{s}=[0.046, 0.14]$ GeV.
Finally, for the ``Fit C" with  effective quark masses $m_{u,d}, m_{s}=[0.14, 0.14]$ GeV, the authors in Ref. \cite{Ahmady} chose the following parameters
\begin{equation}
\!\frac{\sigma_0}{{\rm mb}} = 29.9,\; \gamma_s \!=\! 0.724,\;  \lambda \!=\! 0.206, \;x_0 \!=\! 6.33 \times10^{-6},\end{equation} with a $\chi^2/\mbox{d.o.f}=1.07$, in the case of effective quark masses $m_{u,d}=[0.046, 0.14]$ GeV.

Hence, we can employ  the nuclear  configurational entropy \cite{Karapetyan:2017edu,Karapetyan:2016fai} to \emph{derive} at least one of these parameters used in Ref. \cite{Ahmady}. In fact, different papers concerning various aspects of experimental, phenomenological 
and theoretical observations this scheme has been extensively used with predictions that precisely fit experimental data 
(see \cite{Bernardini:2016hvx, Bernardini:2016qit,Casadio:2016aum,Gleiser:2012tu,Sowinski:2015cfa,Gleiser:2011di,Gleiser:2015rwa} and references therein).
 
 We should first calculate the spatial Fourier transform of the cross sections of the diffractive $\rho$ and $\phi$ meson production \cite{Ahmady}. 
Then, in order to calculate the nuclear configurations, one should use as more appropriate value the cross sections as spatially-localised, square integrable, functions, instead of the energy density. 
As was previously defined in Refs. \cite{Karapetyan:2017edu,Karapetyan:2016fai}, we can determine the nuclear configurational entropy as the Fourier transform of the cross section
\begin{equation}
\label{34}
{\boldsymbol\sigma} ({ \bf k})=\frac{1}{2\pi}\! \int_{-\infty}^\infty\int_{-\infty}^\infty  \!\sigma({\bf r})\,\exp({i{\bf k} \cdot{\bf r}}) d {\bf r}\,\,,
\end{equation}  with the respective modal fraction,
\begin{equation}\label{modall}
f_{\boldsymbol\sigma}({\bf k})=\frac{\vert\,\boldsymbol\sigma({\bf k})\,\vert^2}{\int_{-\infty}^\infty\int_{-\infty}^\infty\vert\,{\boldsymbol\sigma({\bf k})\vert\,^2}\,d{\bf k}}.
\end{equation}  
 Finally, the nuclear configurational entropy is calculated by  \cite{Gleiser:2012tu}:
\begin{equation}
\label{333}
S_c\left[f_{\boldsymbol\sigma}\right] \,= \, - \int_{-\infty}^\infty\int_{-\infty}^\infty f_{\boldsymbol\sigma}({\bf k} ) \log  f_{\boldsymbol\sigma} ({\bf k})  d {\bf k}. 
\end{equation} 
 
Now Eqs. (\ref{34} - \ref{333}) for the nuclear configurational entropy can be computed upon the $\rho$ and $\phi$ mesons cross sections.
According to the Ref. \cite{Ahmady}, the best fit for the $\rho$ vector meson production was  the set of adjustive parameters labeled as ``Fit A", and in the case of $\phi$ production -- ``Fit B", thus we decided to calculate the nuclear configurational entropy exactly for above mentioned set of parameters by fixing three of them in each set, and let free just the polarisation parameter $\lambda$.
\begin{figure}[!htb]
       \centering
                \includegraphics[width=0.9\linewidth]{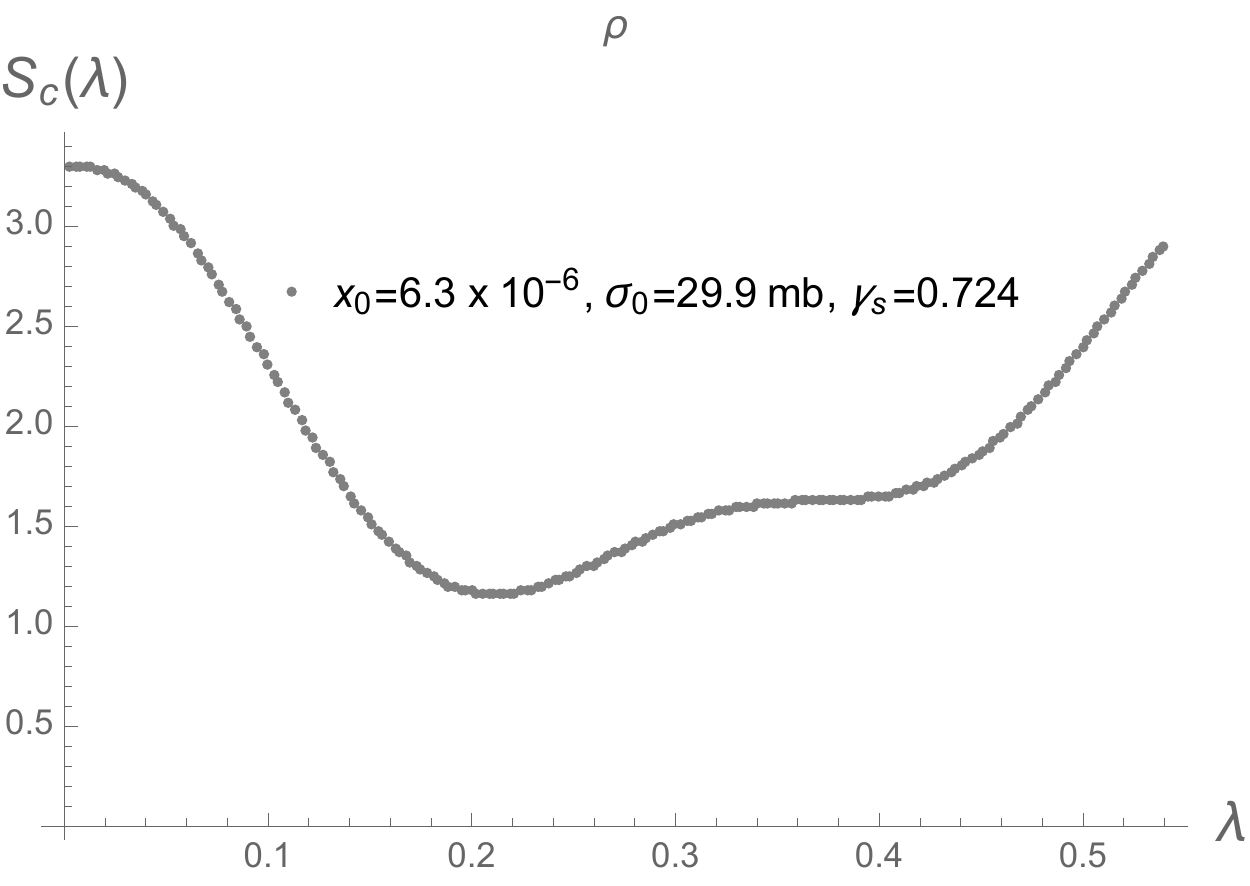}
                \caption{Nuclear configurational entropy for the $\rho$ meson production, as a function of the $\lambda$ parameter, in the   Color Glass Condensate. The  $\lambda$ adjustive polarisation parameter in the case of best fit for $\rho$ meson production cross section
 at three fixed parameters, $\sigma_0, x_0$ and $\gamma_s$ \cite{Ahmady}.}
                \label{curv-ricci}
\end{figure}

\begin{figure}[!htb]
       \centering
                \includegraphics[width=0.9\linewidth]{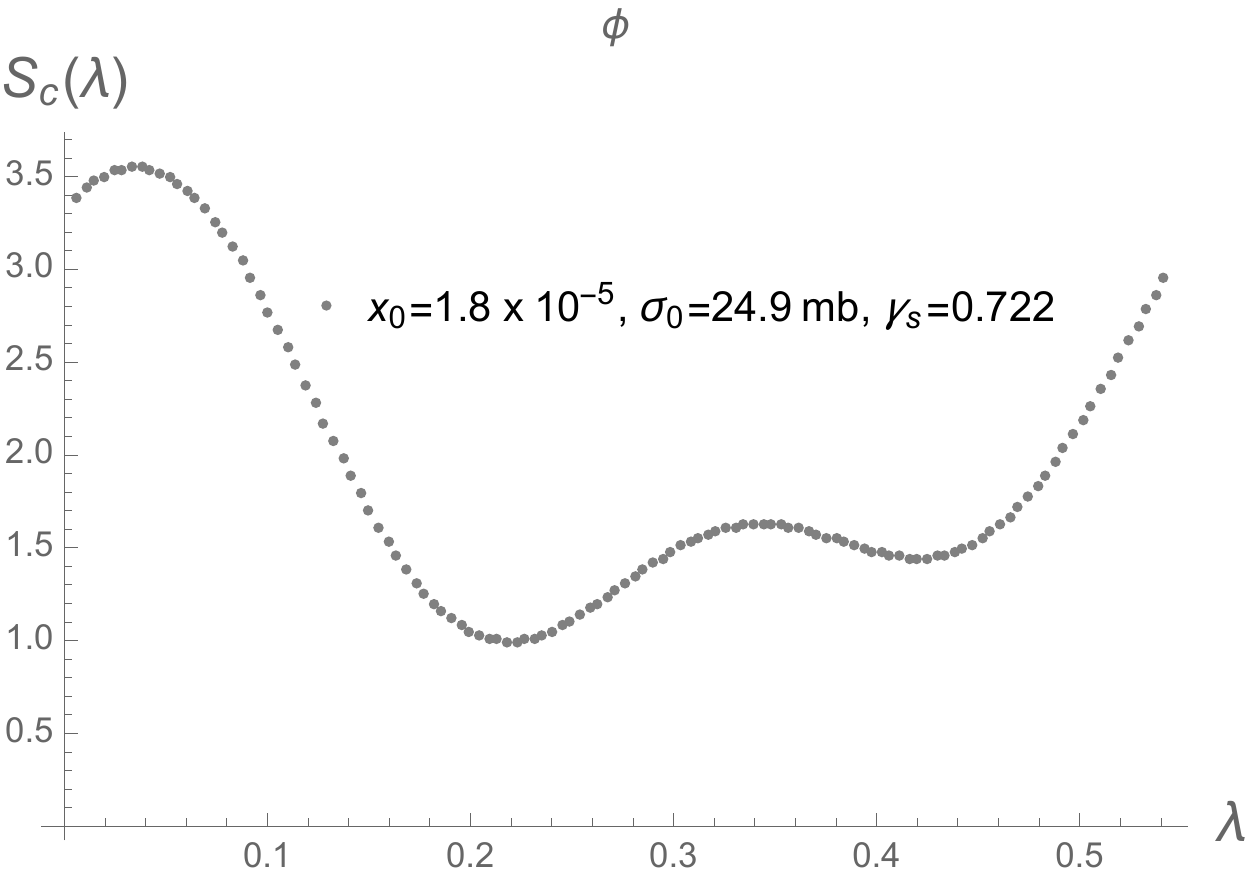}
                \caption{Nuclear configurational entropy for the $\rho$ meson production, as a function of the $\lambda$ parameter, in the   Color Glass Condensate. The  $\lambda$ adjustive polarisation parameter in the case of best fit for $\phi$ meson production cross section
 at three fixed parameters, $\sigma_0, x_0$ and $\gamma_s$ \cite{Ahmady}.}
                \label{curv-ricci}
\end{figure}

Calculated by Eq. (\ref{333}), the nuclear configurational entropy is then plot in Figs. 1 and 2, for the set of fixed parameters, $\sigma_0, x_0$ and $\gamma_s$ and one adjusted, the polarisation parameter $\lambda$, which corresponds to the best fit predictions for diffractive $\rho$ and $\phi$ production cross sections, respectively, according to Ref. \cite{Ahmady}. The error of the calculation is less than 0.1\% compared with the experimental data \cite{Ahmady}.  
One can see in Fig. 1, that the minimum of the nuclear configurational entropy coincides with the value of $\lambda=0.206$ parameter, adjusted in Ref. \cite{Ahmady} for the case of $\rho$ meson production cross section prediction.
Our numerical calculation consistently confirms the results observed in experiments, which, however,  also have been found
 by simulations of the CGC dipole model parameters using the definitive 2015 HERA data.
The calculations of the nuclear configurational entropy for  the case of $\phi$ meson production cross section show two different minima, first of which and the deepest, at $\lambda=0.222$, correspond to that one which has been predicted by simulation in Ref. \cite{Ahmady}.
The presence of the second minimum can be interpreted as an alternative value of $\lambda$, which also can be used
in order to predict the meson production, however, in the such case, the set of the adjusted parameters will be changed 
and also can be coupling and sensitive to the variation in the quark masses.
In order to provide the natural option of the set of adjusted parameters, we can conclude that the calculation of the minima of the configurational entropy establish the onset of the quantum regime for the prediction of $\rho$ and $\phi$ meson production cross sections, matching experimental data with great precision.

\section{Conclusions}
In the context of the AdS/QCD  and the Color Glass Condensate, the nuclear configurational entropy was used to find the parameters in order to compute the cross sections for diffractive $\rho$ and $\phi$ meson production.
The computation of the nuclear configurational entropy minimal points indicates the prevalence of states with such parameters obtained at the end of the previous section, consisting of a natural choice for the  parameters, which are necessary to reproduce quite adequately the experimental data. 
Such minima are represented by set of three fixed parameters $\sigma_0, x_0$ and $\gamma_s$ \cite{Ahmady}
and one adjustive polarisation parameter,  $\lambda=0.206$,  in the case of $\rho$ meson production cross section. For the $\phi$ diffractive meson production cross section, the minimum of the nuclear configurational entropy corresponds to the polarisation parameter $\lambda=0.222$, that is exactly the same used in Ref. \cite{Ahmady} for fitting, based in HERA data. 
Our calculation reproduces the results of the fitting in Ref. \cite{Ahmady}, where
simulations have been computed for the CGC dipole model parameters, using the definitive 2015 HERA data.
The dipole cross section together with the holographic light-front meson wavefunction are  successful methods to describe simultaneously diffractive $\rho$ and $\phi$ production with a set of light quark masses, and the nuclear configurational entropy plays a prominent role on deriving the polarisation parameter $\lambda$ that minimizes it. 
A direction to be further studied encompasses quantum mechanics fluctuations. The wave function used 
in those equations can be explored in the context of topological defects proposed in Refs. \cite{Bazeia:2013usa,Bazeia:2012qh,Bernardini:2016rgb,Bernardini:2012bh}, in the configurational entropy setup \cite{Correa:2016pgr,Correa:2015vka}.
 
  \acknowledgements
GK thanks to FAPESP (grant No. 2016/18902-9), for partial financial support.

\begin{thebibliography}{999}
%

\bibitem{Bernardini:2016hvx} A. E. Bernardini and R. da Rocha, Phys.\ Lett.\ B {\bf 762}, 07 (2016).
\bibitem{Bernardini:2016qit} A. E. Bernardini,  N. R. F. Braga and R. da Rocha, Phys.\ Lett.\ B {\bf 76}, 81 (2017).
\bibitem{Casadio:2016aum} R. Casadio and R. da Rocha, Phys.\ Lett.\ B {\bf 763}, 434 (2016).
\bibitem{Braga:2017fsb} 
  N.~R.~F.~Braga and R.~da Rocha,
  Phys.\ Lett.\ B {\bf 776}, 78 (2018).
  \bibitem{Braga:2016wzx} 
  N.~R.~F.~Braga and R.~da Rocha,
  Phys.\ Lett.\ B {\bf 767}, 386 (2017).
  \bibitem{daSilva:2017jay} 
  A.~Goncalves da Silva and R.~da Rocha,
  Phys.\ Lett.\ B {\bf 774}, 98 (2017).
\bibitem{Ma:2018wtw} 
  C.~W.~Ma and Y.~G.~Ma,
  Prog.\ Part.\ Nucl.\ Phys.\  {\bf 99}, 120 (2018).
  \bibitem{Gleiser:2012tu} M. Gleiser and N. Stamatopoulos, Phys.\ Rev.\ D {\bf 86}, 045004 (2012).
\bibitem{Sowinski:2015cfa} M. Gleiser and D. Sowinski, Phys.\ Lett.\ B {\bf 747}, 125 (2015).
\bibitem{Gleiser:2011di} M. Gleiser and N. Stamatopoulos, Phys.\ Lett.\ B{\bf 713}, 304 (2012).
\bibitem{Gleiser:2015rwa} M. Gleiser and N. Jiang, Phys.\ Rev.\ D {\bf 92}, 044046 (2015).
\bibitem{Gleiser:2013mga} 
  M.~Gleiser and D.~Sowinski,
  Phys.\ Lett.\ B {\bf 727}, 272 (2013).
\bibitem{Karapetyan:2017edu} 
  G.~Karapetyan,
  EPL {\bf 118}, no. 3, 38001 (2017).
\bibitem{Karapetyan:2016fai} 
  G.~Karapetyan,
  EPL {\bf 117}, no. 1, 18001 (2017).
\bibitem{Alves:2017ljt} 
  A.~Alves, A.~G.~Dias and R.~Silva,
  Braz.\ J.\ Phys.\  {\bf 47}, no. 4, 426 (2017).
\bibitem{Alves:2016koo} 
  A.~Alves, A.~G.~Dias and K.~Sinha,
  JHEP {\bf 1608}, 060 (2016).

\bibitem{Karch:2006pv}  A.~Karch, E.~Katz, D.~T.~Son and M.~A.~Stephanov,
  Phys.\ Rev.\ D {\bf 74},  015005 (2006).
\bibitem{Huang} Chao Huang, Bo-Qiang Ma, arXiv:1705.06399 [hep-th].
\bibitem{Ahmady} M. Ahmady, R. Sandapen, N. Sharma, Phys.\ Rev.\ D {\bf 94}, 074018 (2016). 
\bibitem{Ma:2015lpa} C. W. Ma, H. L. Wei, S. S. Wang, Y. G. Ma, R. Wada and Y. L. Zhang, Phys.\ Lett.\ B {\bf 742}, 19 (2015).

\bibitem{Brodsky:2006uqa} S. J. Brodsky and G. F. de Teramond, Phys.\ Rev.\ Lett. {\bf 96}, 201601 (2006).

\bibitem{Bazeia:2013usa} 
  D.~Bazeia, R.~Menezes and R.~da Rocha,
  Adv.\ High Energy Phys.\  {\bf 2014}, 276729 (2014).

\bibitem{Bazeia:2012qh} 
  D.~Bazeia, L.~Losano, R.~Menezes and R.~da Rocha,
  Eur.\ Phys.\ J.\ C {\bf 73}, 2499 (2013).



\bibitem{Bernardini:2012bh} 
  A.~E.~Bernardini and R.~da Rocha,
  Adv.\ High Energy Phys.\  {\bf 2013}, 304980 (2013).

\bibitem{Bernardini:2016rgb} A. E. Bernardini and R. da Rocha, Phys.\ Lett.\ A {\bf 380}, 2279 (2016).

\bibitem{Correa:2016pgr} 
  R.~A.~C.~Correa, D.~M.~Dantas, C.~A.~S.~Almeida and R.~da Rocha,
  Phys.\ Lett.\ B {\bf 755}, 358 (2016).


\bibitem{Correa:2015vka} 
  R.~A.~C.~Correa and R.~da Rocha,
  Eur.\ Phys.\ J.\ C {\bf 75}, no. 11, 522 (2015).
\end{thebibliography}
\end{document}